\documentclass[manuscript]{emulateapj}
\usepackage{natbib}
\usepackage{graphicx}

\newcommand{\chandra}{{Chandra}}
\newcommand{\spitzer}{{Spitzer}}
\newcommand{\hst}{{Hubble}}
\newcommand{\vla}{{VLA}}
\newcommand{\galex}{{GALEX}}

\newcommand{\um}{~$\mu$m}

\newcommand{\hubble}{$H_0=70$ km s$^{-1}$ Mpc$^{-1}$}
\newcommand{\etal}{{\em et al.}}
\bibpunct{(}{)}{,}{a}{}{}

\shorttitle{3C 31 IR jet} 
\shortauthors{Lanz et al.}

\begin{document}

\title{The Infrared Jet in 3C 31}
\author{Lauranne Lanz\altaffilmark{1}, Amelia Bliss\altaffilmark{2}, 
Ralph P.~Kraft\altaffilmark{1}, Mark Birkinshaw\altaffilmark{2}, Dharam V. Lal\altaffilmark{1}, 
William~R.~Forman\altaffilmark{1}, Christine~Jones\altaffilmark{1}, 
Diana M. Worrall\altaffilmark{2}
}
\altaffiltext{1}{Harvard-Smithsonian Center for Astrophysics, 60 Garden St., 
Cambridge, MA 02138, USA; llanz@head.cfa.harvard.edu}
\altaffiltext{2}{H H Wills Physics Laboratory, University of Bristol, Tyndall
Avenue, Bristol BS8 1TL, UK}

\begin{abstract}
We report the detection of infrared emission from the jet of the nearby FR I radio
galaxy 3C 31.  The jet was detected with the IRAC instrument on Spitzer at 4.5\um, 
5.8\um, and 8.0\um~out to 30$\arcsec$ (13 kpc) from the nucleus. We measure radio, 
infrared, optical, and X-ray fluxes in three regions along the jet determined by 
the infrared and X-ray morphology. Radio through X-ray spectra in these regions 
demonstrate that the emission can be interpreted as synchrotron emission from a 
broken power-law distribution of electron energies. We find significant differences 
in the high energy spectra with increasing distance from the nucleus. 
Specifically, the high energy slope increases from 0.86 to 1.72
from 1 kpc to 12 kpc along the jet, and the spectral break likewise increases in frequency
along the jet from 10-100's of GHz to $\sim$20 THz.
Thus the ratio of IR to X-ray flux in the jet increases by at least an order of magnitude with 
increasing distance from the nucleus.  We argue that these changes cannot simply be the 
result of spectral aging and that there is ongoing particle acceleration through this region 
of the jet.  The effects of mass loading, turbulence, and jet deceleration, however these 
processes modify the jet flow in detail, must be causing a change in the electron energy 
distribution and the efficiency of particle acceleration.
\end{abstract}


\section{Introduction}

Extended radio emission in galaxies is the observable manifestation 
of powerful, collimated outflows from supermassive black holes. Only 
$\sim$10\% of active galaxies are radio loud and contain well-formed 
jets, but in all but the most luminous systems, the mechanical power 
of the jets generally dwarfs any radiative losses from an accretion 
disk \citep{bir04}. It is now well established that jets play an 
important role in feedback between the central supermassive black 
hole (SMBH) and its larger scale environment. However, in spite of 
several decades of research, both observational and theoretical, 
basic issues related to jet physics, including particle content, 
velocity structure, the role of magnetic fields, and a host of other 
problems, remain unresolved. Extragalactic jets in low-power 
Fanaroff-Riley type I (FR I) sources are generally only observable 
at frequencies above the radio band via the synchrotron radiation 
from ultra-relativistic particles accelerated at shocks or other 
disturbances in the flow, although at X-ray energies, inverse 
Compton emission can sometimes be important. The significant numbers 
of detections of FR I jets in the IR, optical, and X-ray regimes 
has given us new insights into jet dynamics and particle acceleration 
\citep[see e.g.][for a review]{wor09}.

The radio, optical, and X-ray emission from FR I jets can 
normally be modeled as synchrotron emission from a single electron 
energy distribution described by a broken power law.  Electrons emitting 
synchrotron radiation have lifetimes that vary inversely with
their energy, so multi-wavelength observations are expected to show
that higher-energy electrons are concentrated closer to the sites of
particle acceleration than their lower-energy counterparts.  In typical 
equipartition magnetic fields of FR I radio galaxies, the lifetimes
of IR/optical and X-ray emitting particles are $\sim$10$^4$ and 10$^2$ yrs,
respectively \citep[e.g.][]{wor09}.  Light travel times for kiloparsec jets 
demonstrate that the ultra-relativistic particles must be accelerated 
in situ.  Proposed acceleration mechanisms include magnetohydrodynamic 
turbulence and shock acceleration, among others \citep[e.g.][]{wor09}.

3C 31 is a nearby twin-jet radio source hosted within the early-type 
galaxy NGC 383, the dominant member of a small group of galaxies that 
also includes the nearby companion NGC 382. At a redshift of 0.0168 
($D_{L}$=72.9 Mpc) \citep{dev91}\footnote{We adopt a cosmology with \hubble, 
$\Omega_{M}=0.27$, and $\Omega_{\Lambda}=0.73$, resulting in a linear 
projected scale of $0.34$ kpc per arcsec at the source. We deproject
distances assuming the jet is 52$^{\circ}$ from the line of sight
\citep{lai02b}.}, 3C 31 was one 
of the earliest FR I galaxies to be studied in depth in the radio 
\citep{bur77,fom80} and remains a canonical example of this class. The 
twin jets are aligned roughly along a north-south axis.  The northern 
jet is significantly brighter at GHz frequencies 
in the central 10 kpc \citep{lara97}, suggesting, that in this regime at least, the flow is moderately relativistic 
and this jet is oriented towards us \citep{lai08}. The northern jet is straight 
and narrow within 45$''$ (19 kpc) of the nucleus, but beyond this distance, flares 
dramatically suggesting that it is losing kinetic energy \citep{bur77}. 
On larger scales, both the forward and reverse jets extend at least 
20$\arcmin$ (500 kpc) from the nucleus \citep{lai08}.  The radio morphology of the jet 
becomes wavy and diffuse at these distances.  Subsonic motions of the 
external medium are probably important in determining the large scale 
radio morphology of the jet \citep[e.g.][]{loken95}. The inner 8.5 kpc of the northern jet has been 
detected in the optical \citep{cro03, but80}, and the inner 2.7 kpc in 
the X-ray \citep{har02}. It has also previously been detected at lower
spatial resolution in the infrared \citep{tan97}. \citet{har02} modeled the jet emission as 
synchrotron radiation from a population of ultra-relativistic electrons. 

The relatively regular two-sided nature of 3C 31 in radio has permitted detailed
kinematic modeling of the jets. In particular, Laing and Bridle (2002a, 
2002b, 2004) created a 2-dimensional model of the jet flow to determine 
the velocity profile as a function of distance down the jet axis.  
Laing and Bridle (2002a, 2004) extended their work to include dynamical 
modeling by incorporating the pressure of the exterior gas as a function 
of distance based on X-ray measurements. They divided the inner jet into 
three regions: an inner region (extending to $2\farcs5$ (1.1 kpc)) wherein 
the jet expands as a cone with a half-opening angle of $8\fdg5$, a flaring 
region ($2\farcs5-8\farcs3$; 1.1 - 3.6 kpc) with gradually larger opening
angle, and an outer region ($8\farcs3-28\farcs3$; 3.6 - 12.2 kpc) wherein the jet once 
more expands conically, but with a half-opening angle of $16\fdg75$. They 
found that these regions have distinct kinematic properties and that 
the jet deceleration is consistent with the entrainment of gas from the
X-ray emitting medium.

In this paper, we present a detection of infrared emission from the jet 
of 3C 31 as imaged by the Spitzer Space Telescope Infrared Array 
Camera (IRAC). \spitzer~has detected infrared jets in a few nearby 
galaxies, including M87 \citep{shi07, for07} and Cen A (NGC 5128) \citep{har06, bro06}.
We discuss 3C 31's inner jet morphology in the infrared, X-ray, and radio. 
We measure jet IR flux densities in the IRAC and MIPS bands in the flaring 
and outer regions defined by \citet{lai02a}.  We use the resulting energy 
distribution for each region to constrain particle acceleration in the jet. 
Images have North to the top and East to the left, angles are given
counterclockwise from North, and spectral indices $\alpha$ are defined 
such that flux density at frequency $\nu$~$\propto~\nu^{-\alpha}$. We use the $52^{\circ}$
angle to the line of sight for the jet determined by \citet{lai02b} (hereafter L02) to give 
linear distances as deprojected distances.

\begin{figure*}
\centerline{\includegraphics[width=0.55\linewidth]{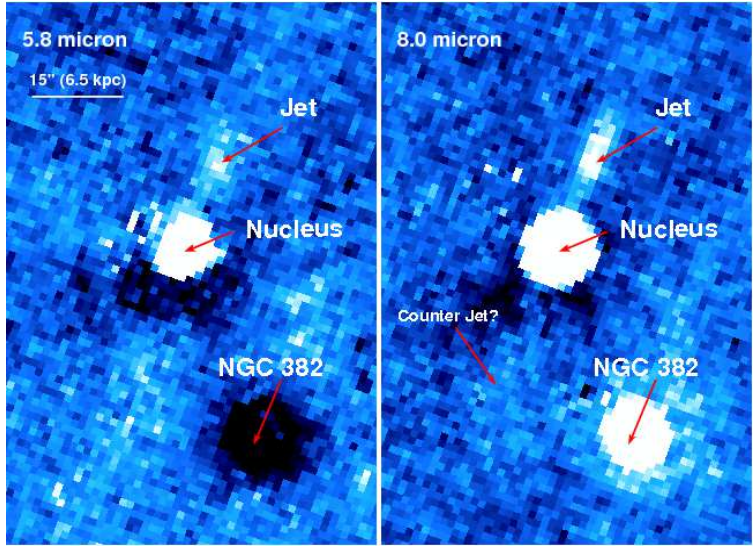}}
\caption{Spitzer IRAC images of the nonstellar emission at 5.8\um~(left) 
and 8.0\um~(right) in a $1\farcm5\times1\arcmin$
field around the nucleus. The stellar
emission was modeled using the 3.6\um~scaled to
compensate for zero point offsets and aperture
corrections. The 3.6\um~emission was color-corrected using
a color profile determined for each of the longer IRAC bands. 
Jet emission is seen in  the region to the north-northwest 
of the nucleus, both of which are indicated by the red arrows.
The technique used to subtract the stellar emission of NGC 383
results in different residuals of NGC 382 but does not impact
our study of NGC383's jet. The scale bar is 15$\arcsec$ (6.5~deprojected kpc).
\label{nons}}
\end{figure*}

\begin{figure*}
\centerline{\includegraphics[width=0.55\linewidth]{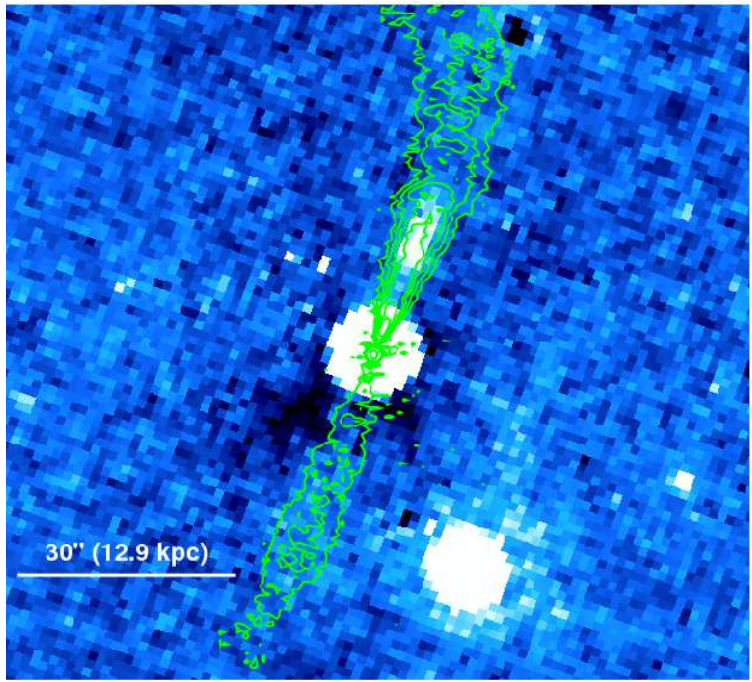}}
\caption{Spitzer 8.0\um~image of the jet of 3C 31 in 1$\farcs2$ pixels with 
8.4 GHz radio contours overlaid, showing the relative location of 
the IR emission in the larger radio jet.
The image was smoothed with a Gaussian $0\farcs6$ wide after the stellar
emission was subtracted. The 8.4 GHz contours are at 0.1, 0.2, 0.4, 0.6, 
1.2, 2, 20 mJy beam$^{-1}$ ($0\farcs69\times0\farcs67$ beam).
\label{irradovl}}
\end{figure*}

\section{Observations and Data Reduction}

\subsection{\spitzer~Data}

IRAC \citep{faz04} observations of 3C 31 were obtained on 2006 February 
15 and 22 (\spitzer~PID 3418) at 3.6\um, 4.5\um, 5.8\um, and 8.0\um. Each 
observation consisted of 24 frames of 30~s integrations covering the galaxy and 
the nearby field obtained using a medium cycling dither. For our analysis, 
we retrieved the Basic Calibrated Data 
(BCD) version S14.0 pipeline products from the \spitzer~archive. For the 
8.0\um~observation, light scattered from the galaxy nucleus along the 
detector-array rows was fit and subtracted from the BCD frames using 
custom software as in \citet{ash09}. We then coadded the 30~s BCD 
frames using version 4.1.2 of {\tt IRACproc} \citep{sch06} 
into mosaics having 1\farcs2 pixels, i.e., the native IRAC pixel size. A 
$4\farcm8\times4\farcm8$ region centered on the galaxy was selected for 
analysis in each IRAC mosaic. 

\begin{figure*}
\centerline{\includegraphics[width=0.6\linewidth]{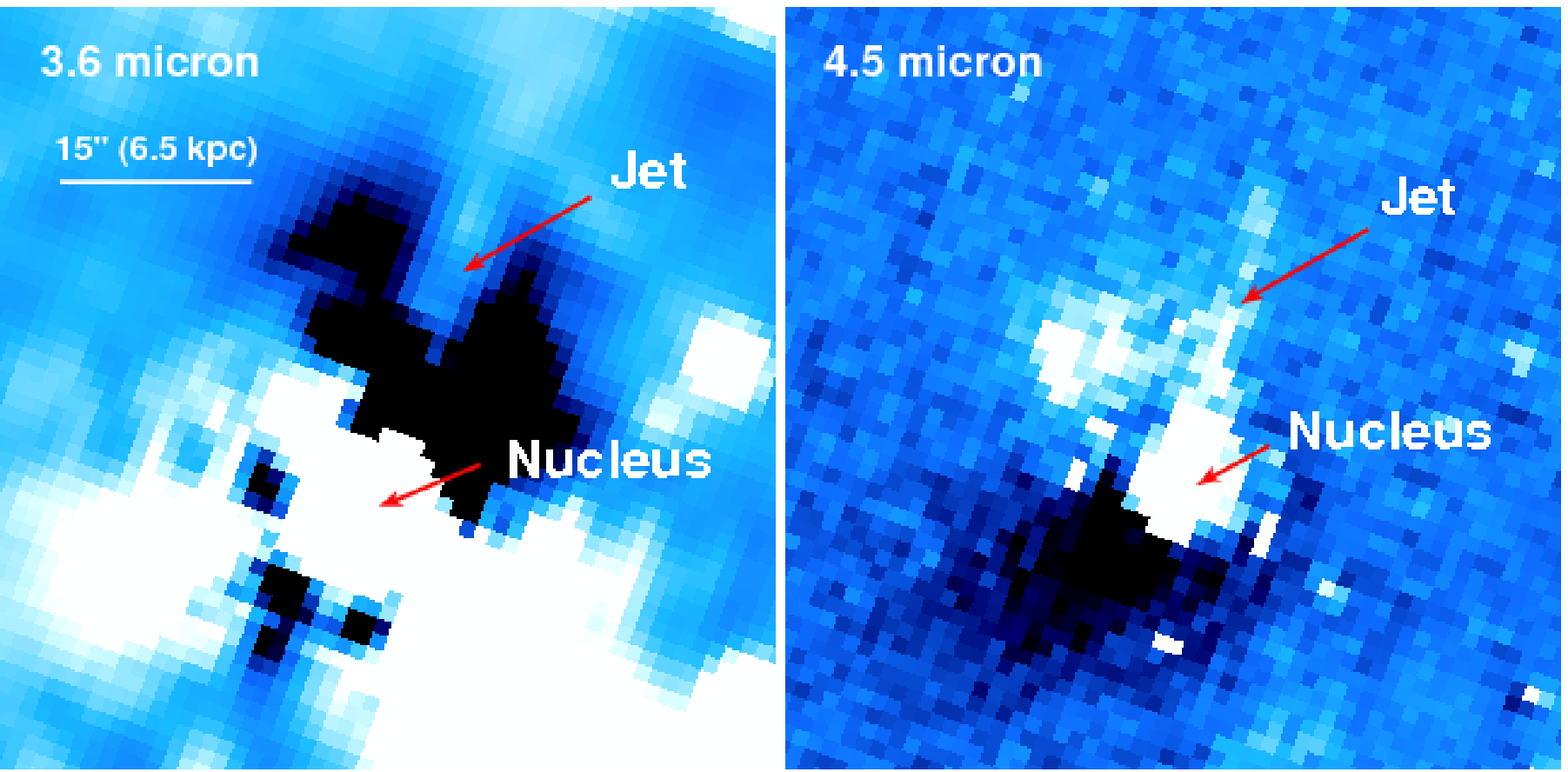}}
\caption{Spitzer IRAC images at 3.6\um~(left) and 4.5\um~(right)
in a square arcminute field around the nucleus. The 3.6\um~image
has had an azimuthally smoothed version of the image subtracted, 
in order to emphasize the jet emission. The 4.5\um~image is of
the nonstellar emission where the stellar emission was modeled
based on the scaled 3.6\um~emission in the same manner as for
Figure \ref{nons}.
\label{ch12non}}
\end{figure*}

To detect and determine the jet's infrared morphology, we needed to 
subtract the bright infrared stellar emission. We modeled the stellar emission 
using the 3.6\um~emission, as has been done for other galaxies 
\citep[e.g.][]{pah04,tan09}. After subtracting a constant background 
determined in the frames far from NGC 383 and NGC 382, we used GALFIT 
\citep{pen02} to fit the ellipticity and position angle of the galaxy. 
The 3.6\um~image was scaled to correct for differing instrumental parameters 
between the different IRAC bands, specifically different aperture correction 
factors for an infinite aperture and the different photometric 
zeropoints\footnote{IRAC Instrument Handbook Tables 4.1 and 4.7: \\
http://ssc.spitzer.caltech.edu/irac/iracinstrumenthandbook/}.
We then performed photometry using elliptical annuli on the 4.5\um, 5.8\um, and 
8.0\um~background-subtracted images and the associated scaled 3.6\um~images. 
We measured the flux in each annulus and used these to calculate the colors 
between 3.6\um~and each of the longer IRAC bands in each annulus, resulting
in color profiles for NGC 383. We fit these profiles in the annuli from 
$6\farcs5$ to $118\farcs5$ centered on the nucleus. When extrapolated to the galaxy core, 
the fits were used to color-correct the scaled 3.6\um~images to be the 
stellar models at 4.5\um, 5.8\um, and 8.0\um. This method does not take into account
differences in colors due to central features, so we expect that some residual
emission may be present from either the AGN or the dust disk described by \citet{mar00}.
We detected nonstellar jet emission in the 4.5\um, 5.8\um, and 8.0\um~bands, after 
the stellar model was subtracted. The residual images at 5.8\um~and 8.0\um~are shown 
in Figure \ref{nons}. The 8.0\um~image is also shown in Figure \ref{irradovl}
with contours of radio emission in order to localize the infrared emission within the 
larger radio jet. Figure \ref{ch12non} shows the 4.5\um~residual image along with 
a 3.6\um~image, where an azimuthally smoothed version of that image has been subtracted
in order to detect the jet. While the jet is less clearly present at 4.5\um~than in the two longer
IRAC bands, these residual images in the two shorter IRAC bands contain emission 
at the location of the jet. The technique used to subtract the stellar emission of 
NGC 383 results in different residuals for NGC 382, but does not impact our study of the jet.  We also modeled the stellar emission using 
2MASS K-band emission. Finally, we used the IRAF ELLIPSE package 
to model the galaxy emission (Bliss \etal, in preparation). The resulting nonstellar 
images show similar morphology to those in Figures \ref{nons}-\ref{ch12non}. 

We also obtained a mosaic at 24\um~taken with the MIPS instrument 
\citep{rie04} on 2004 December 26 (\spitzer~PID 82). We used GALFIT 
to fit the emission with a galaxy model along with two point 
sources for the nuclear emission of 3C 31 and for the companion 
galaxy NGC 382. 

\subsection{\chandra, \galex, \hst, 2MASS, and \vla~Data}

3C 31 was observed for 44.41~ks on 2000 November 6 (ObsID 2147)
with the \chandra~Advanced CCD Imaging Spectrometer \citep[ACIS;][]{wei00}.
This observation was previously analyzed by \citet{har02}. GALEX observed 
3C 31 on 2006 November 9 (AIS 42) for 109 seconds in both the NUV and FUV 
filter \citep{mar05}. We obtained the intensity maps in these filters from the 
Multimission Archive at STScI (MAST). We retrieved mosaics of 3C 31 taken by
the Wide Field Planetary Camera 2 (WFPC2) \citep{hol95} on the Hubble
Space Telescope (HST) from the Hubble Legacy archives. 3C 31 was observed through
the F555W and F814W filters on the PC1 chip for 460 seconds each on 1998
September 23 and through the FR680N filter on the WF2 chip for 600 seconds
on 1995 September 01. 2MASS All-Sky Survey Atlas images through all three filters
were obtained from the NASA/IPAC Infrared Science Archive (IRSA)\footnote{
http://irsa.ipac.caltech.edu/applications/2MASS/IM/}. The $VLA$ data we use 
here are observations for 6.5 hrs at 8.4~GHz on 1994 June 14 in the B 
configuration and for 25.5 minutes at 1.4~GHz on 1999 July 18 in the A 
configuration. We used the Astronomical Image Processing System (AIPS; 
version 31DEC09) to generate maps with beam sizes of 
$0\farcs69\times0\farcs67$ at 8.4~GHz and $1\farcs76\times1\farcs40$
at 1.4~GHz. 

\begin{figure*}
\centerline{\includegraphics[width=0.8\linewidth]{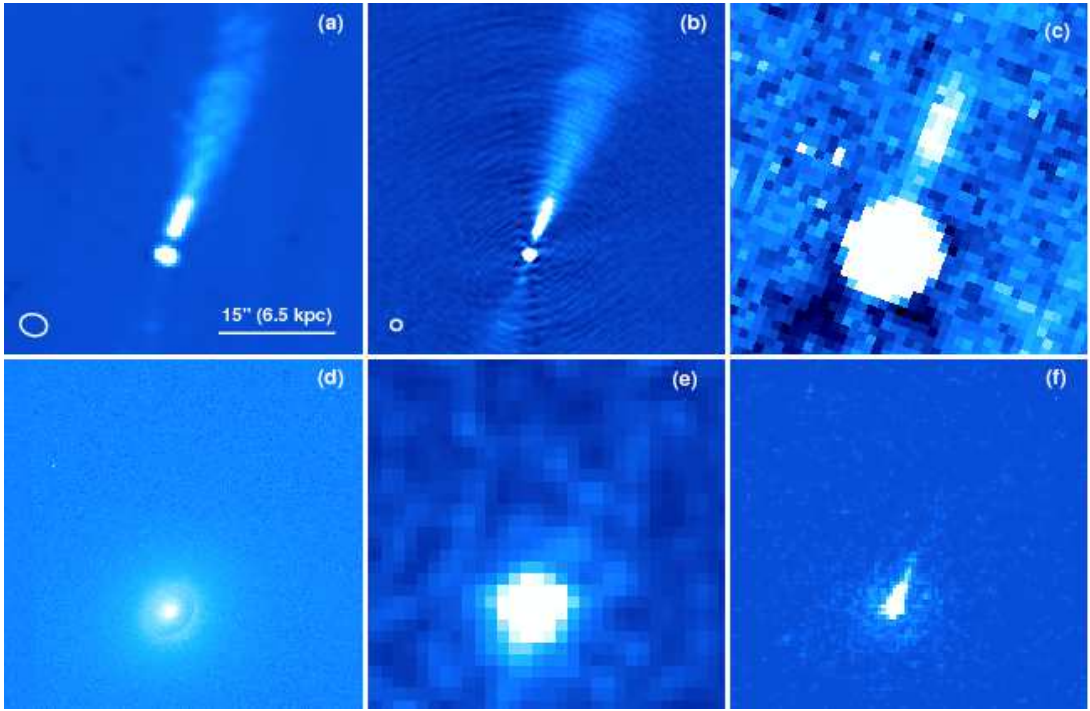}}
\caption{Jet emission at 1.4 GHz (a), 8.4 GHz (b), 
nonstellar 8.0\um~(c), and broadband X-ray energies (0.3-8.0~keV) 
(f). Optical (FR680N, WFPC2 on HST; shown in log scale) and NUV (GALEX) images of the same region are 
shown in (d) and (e) respectively. The 15$\arcsec$ scale 
corresponds to 6.5~kpc.
\label{multi}}
\end{figure*}

\section{Results}

\subsection{Jet Morphology}

\begin{figure}
\centerline{\includegraphics[width=\linewidth]{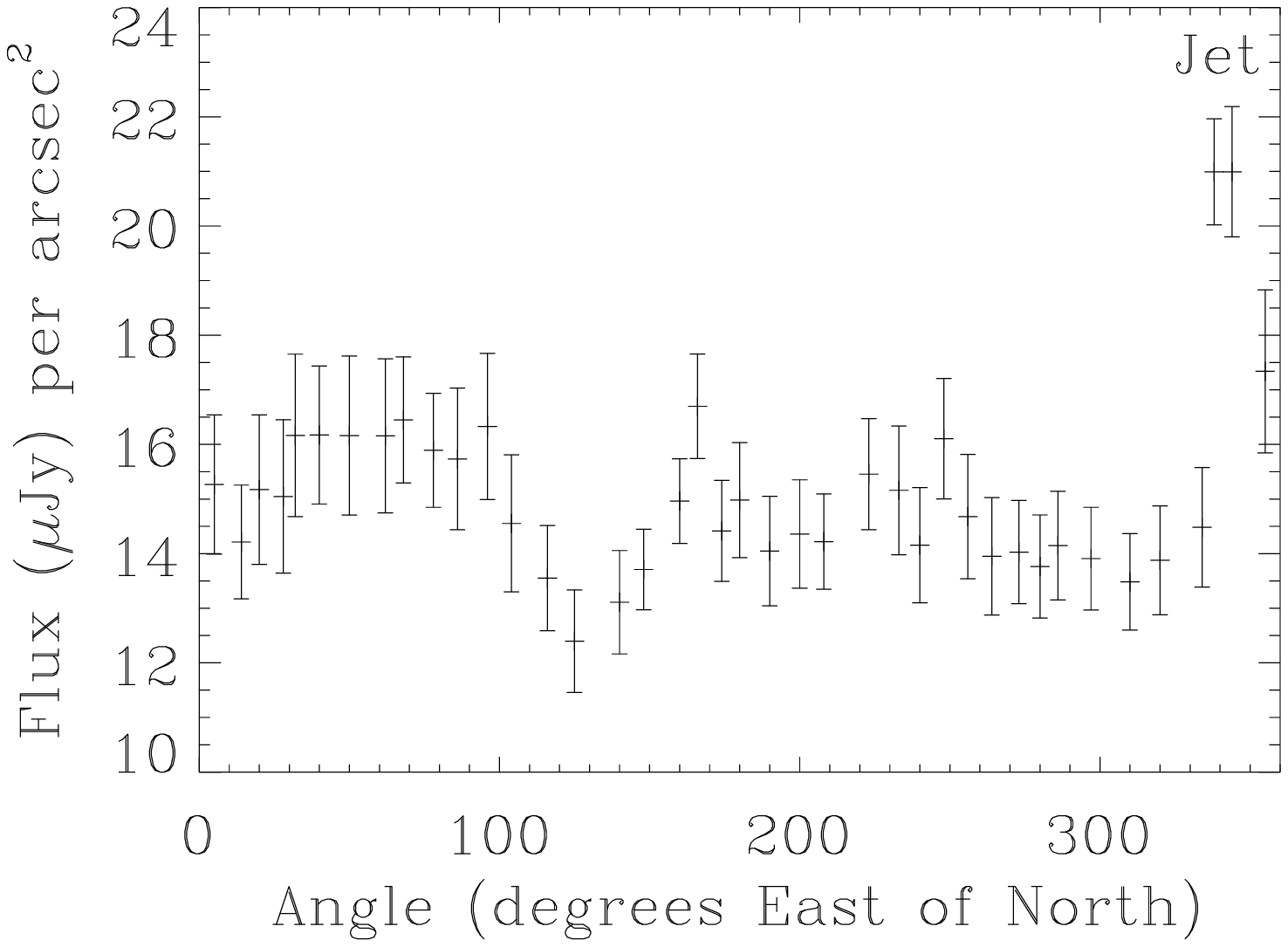}}
\caption{Azimuthal profile of the total 8.0\um~emission 
for an elliptical annulus with inner and 
outer radii of 12$\arcsec$ (5.2~kpc) and 20$\arcsec$ (8.6~kpc) 
with an ellipticity of 0.84. Angles are given counterclockwise 
from north. The jet is located at $330^{\circ}-350^{\circ}$. 
\label{azprof}}
\end{figure}

The infrared jet emission detected at 5.8\um~and 8.0\um~is shown
in Figure \ref{nons}. The jet and nucleus are identified with
red arrows. The northern jet is detected from 8$\arcsec$ (3.5~kpc) 
to 24$\arcsec$ (10.3~kpc), with a position angle of  $-25^{\circ}$. 
The X-ray emission from the northern radio jet has previously been 
described by \citet{har02}. It extends approximately 8$\arcsec$ 
(3.5~kpc) from the nucleus. We have a 3.5$\sigma$ detection of X-ray 
emission from 8$\farcs5$ (3.5~kpc) to 15$\arcsec$ (6.5~kpc). 
\citet{bur77} mapped 3C 31 at 5 radio frequencies between 
0.4~GHz and 15~GHz. More recently, \citet{lai08} obtained high 
resolution \vla~images from 1365-8440~MHz. 
An infrared image of the jet with 8.4 GHz contours of the inner radio 
emission is shown in Figure~\ref{irradovl}. The radio 
emission shows a pair of jets, whose width increases with distance 
from the nucleus. The northern jet is brighter than the southern jet 
by about an order of magnitude between $2\farcs5-6\arcsec$ 
(1.1-2.6~kpc) and remains straight for $\sim$45$\arcsec$ before
gently bending to the north, while the southern jet extends 
$120\arcsec$ before exhibiting a strong bend. The jet is not detected 
in the \hst~or \galex~observations. \citet{cro03} detected jet 
emission using the 4 m Mayall Telescope in the B and R bands. Images 
of the jet emission in the radio, IR, and X-ray, as well as the 
corresponding region in the optical (with HST) and UV (with GALEX), 
are shown in Figure \ref{multi}. 

To quantify the significance of the detection of the  
northern jet in the infrared, we measured the azimuthal 
surface brightness profile in an elliptical annular sector with an 
ellipticity of 0.87 and with inner and outer semi-major axes of 
12$\arcsec$ (5.2~kpc) and 20$\arcsec$ (8.6~kpc), where the peak of 
the infrared jet emission is located based on the residual images shown 
in Figure \ref{nons}. The total 8.0\um~surface brightness in this 
annulus is shown in Figure \ref{azprof}. The statistically significant 
surface brightness enhancement between 330$^{\circ}$ and 350$^{\circ}$ 
is due to the northern jet emission.

In contrast to the northern jet, the counterjet is only 
marginally detected in the two longest wavelength IRAC bands. 
Faint IR emission is noted in the Spitzer image shown in 
Figure \ref{irradovl}, but the statistical significance 
of this emission is low (less than 3$\sigma$). The point estimate 
of the flux density of the counterjet is sensitive to the details of the model
used to subtract the stellar emission, but the statistical
significance of the detection is insensitive to the stellar model.
We measured the flux in the IRAC bands in a 
$10\arcsec\times5\arcsec$ (4.3 kpc$\times$2.2 kpc) aperture 
beginning 15$\arcsec$ from the nucleus centered in the radio contours. 
We find flux densities with 1$\sigma$ uncertainties of 
$35\pm15$ $\mu$Jy at 5.8\um~and $16\pm16$ $\mu$Jy at 8.0\um. In
comparison, the forward jet flux density at the same distance
from the nucleus is significantly greater at $85\pm16$ $\mu$Jy 
at 5.8\um~and $142\pm10$ $\mu$Jy at 8.0\um.

\begin{figure}
\centerline{\includegraphics[width=\linewidth]{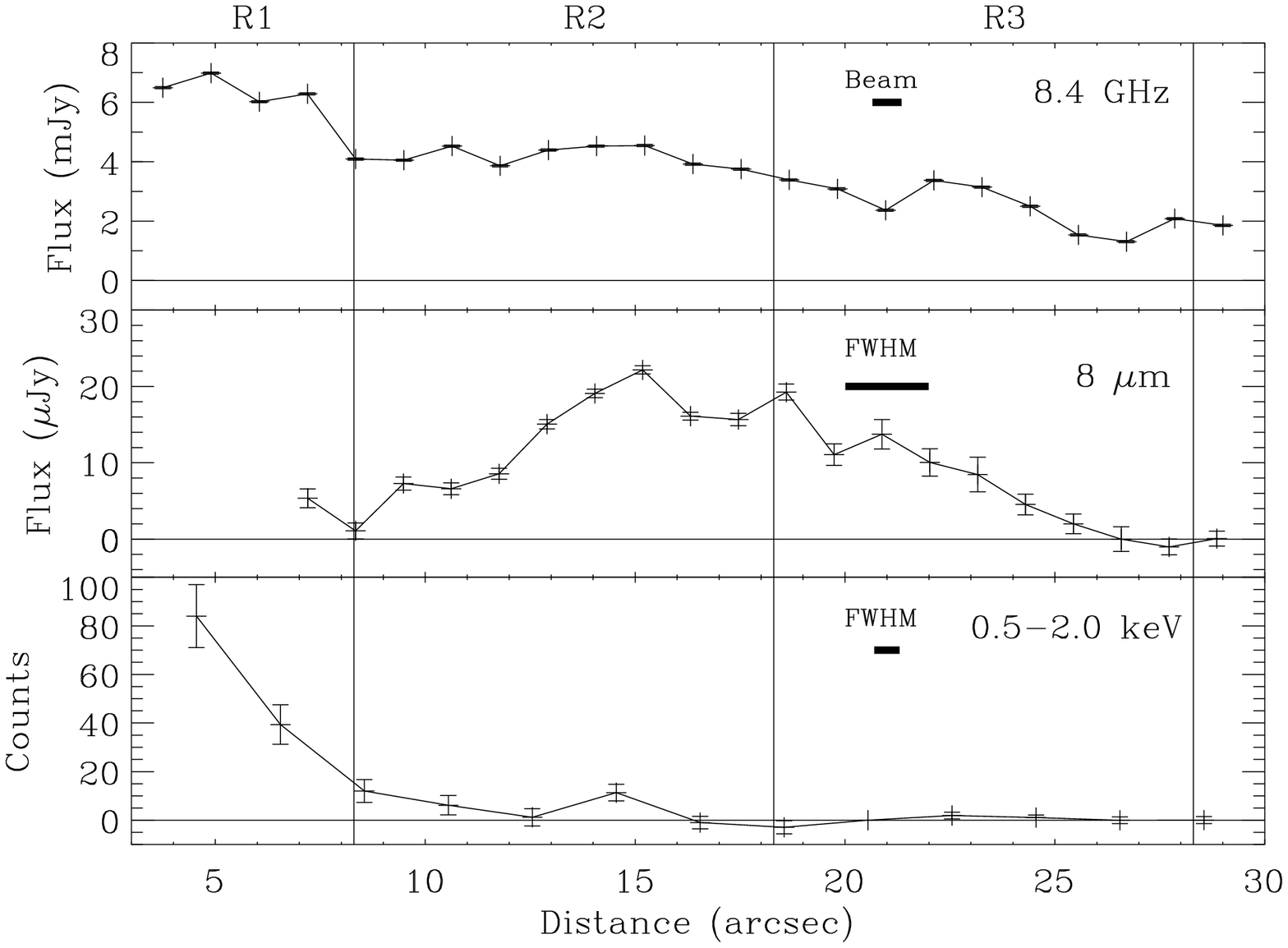}}
\caption{Longitudinal profile of the northern jet emission at 8.4~GHz, 
8.0\um~and 0.5-2.0~keV. Flux densities were measured in 
rectangles 4$\arcsec$ wide and $1\farcs2$ long along 
the jet axis for the infrared and radio. The 8.0\um~emission 
was taken from the nonstellar image (Fig. \ref{nons}). X-ray flux densities were
measured in rectangles 4$\arcsec$ wide and 2$\arcsec$ long. We removed
the X-ray emission from hot ambient gas, determined
in similar rectangles adjacent to the jet. The regions that will
be used to extract spectra are marked at the top of the plot.
\label{longnon}}
\end{figure}

Figure \ref{longnon} shows the longitudinal profile of the northern jet emission at 
8.4~GHz, 8.0\um, and in the soft X-ray band (0.5-2.0~keV). In measuring
the X-ray profile, we removed the emission from 
the ambient hot gas by subtracting a profile in 
similar rectangles adjacent to the jet. We do not try to deconvolve 
the \spitzer~emission within $7\arcsec$ of the nucleus, which contains 
the dusty disk seen in the optical by \citet{mar00}, as well as 
stellar emission and an unresolved point source. The profiles in
Figure \ref{longnon} were measured in rectangles 
4$\arcsec$ wide and either 1$\farcs2$ (IR and radio) or 2$\arcsec$ (X-ray) 
long along the jet axis. The different bands have different beam sizes, 
which results in an oversampled profile for the Spitzer 8.0\um~band. We
binned the data in fewer regions in the X-ray band to emphasize 
the appearance of this emission. 
The 8.0\um~profile shows that the infrared jet is brightest between 12$\arcsec$ (5.2~kpc) 
and 20$\arcsec$ (8.6~kpc) from the nucleus. \citet{cro03} measured the 
longitudinal profile of the jet in the B band (their Figure 5). They 
detected optical emission from the jet within $\sim25\arcsec$ of the 
nucleus with an increase in brightness coincident with the brightest 
infrared jet emission that we observe. The radio profile shows a slow 
decrease in brightness with distance along the northern radio jet, 
with a bright region where the X-ray emission is strongest closest to the nucleus. 

\subsection{Jet Spectrum and Electron Energy Distribution}

\begin{figure*}
\centerline{\includegraphics[width=\linewidth]{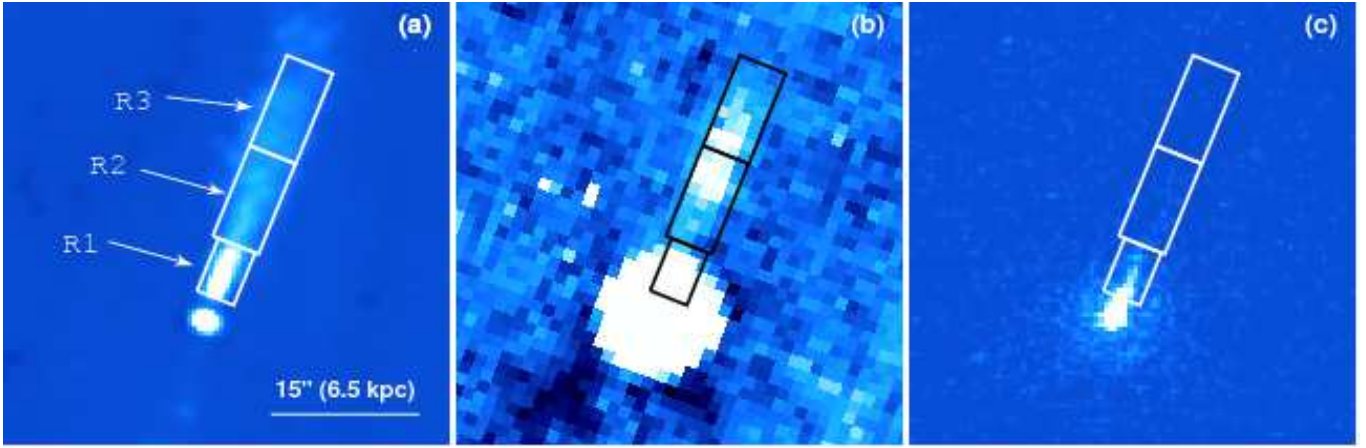}}
\caption{The apertures used for measuring the spectra overlaid
on the 1.4 GHz (a), nonstellar 8.0\um~(b), and broadband X-ray
(0.3-8.0~keV) (c) emission. Note that there is radio emission
in all three apertures and that the IR jet emission is concentrated
in the two regions farthest from the nucleus, while the 
X-ray emission is predominantly in the region closest to the
nucleus. The 15$\arcsec$ scale corresponds to 6.5~kpc.
\label{multiaps}}
\end{figure*}

\begin{deluxetable*}{lccc}
\tabletypesize{\scriptsize}
\tablecaption{Jet Fluxes\label{tflux}}
\tablewidth{0pt}
\tablehead{
\colhead{} & \multicolumn{3}{c}{Flux Density in Region}\\
\colhead{Wavelength} & \colhead{R1} & \colhead{R2} 
& \colhead{R3} \\
\colhead{} & \colhead{($\mu$Jy)} & \colhead{($\mu$Jy)} 
& \colhead{($\mu$Jy)}
}
\startdata
1.4 GHz\tablenotemark{1} & $7.87\times10^4$ 	& $8.98\times10^4$  & $6.73\times10^4$	\\
8.4 GHz\tablenotemark{1} & $3.01\times10^4$ 	& $3.85\times10^4$  & $2.63\times10^4$	\\
24\um & $<960$	   	& $<709$	& $<651$\\
8.0\um & $<1200$ 	& $152\pm25$	& $137\pm9$\\
5.8\um & $<800$        	& $99\pm20.$	& $78\pm19$\\
4.5\um & $<700$         & $<71$     	& $50.\pm7$\\
3.6\um & $<1200$        & $<94$     	& $<48$\\
2.2\um & $<2400$      	& $<570$    	& $<260$\\
1.7\um & $<2500$       	& $<510$     	& $<270$\\
1.2\um & $<1800$       	& $<570$     	& $<380$\\
820.3 nm & $<6.9$	& $<10.$   	& $<10.$\\
667.3 nm & $<17$ 	& $<25$   	& $<25$\\
520.2 nm & $<4.8$	& $<6.9$    	& $<6.9$\\
226.7 nm & $<12$	& $<10.$      	& $<7.5$\\
151.6 nm & $<19$ 	& $<10.$       & $<8.4$\\
0.3-8.0 keV & $(3.3\pm0.3)\times10^{-3}$  & $(3.1\pm1.0)\times10^{-4}$	& $<1.3\times10^{-4}$
\enddata	
\tablecomments{The R1 region extends $2\farcs5-8\farcs3$ and is 
$4\arcsec$ wide. The R2 region extends $8\farcs3-18\farcs3$ and is 
$5\arcsec$ wide. The R3 region extends $18\farcs3-28\farcs3$ and is 
$5\arcsec$ wide. The upper limits are 3$\sigma$ upper limits.}
\tablenotetext{1}{The formal statistical uncertainties are smaller than the significant digits.}
\end{deluxetable*}

\begin{figure}
\centerline{\includegraphics[width=0.9\linewidth]{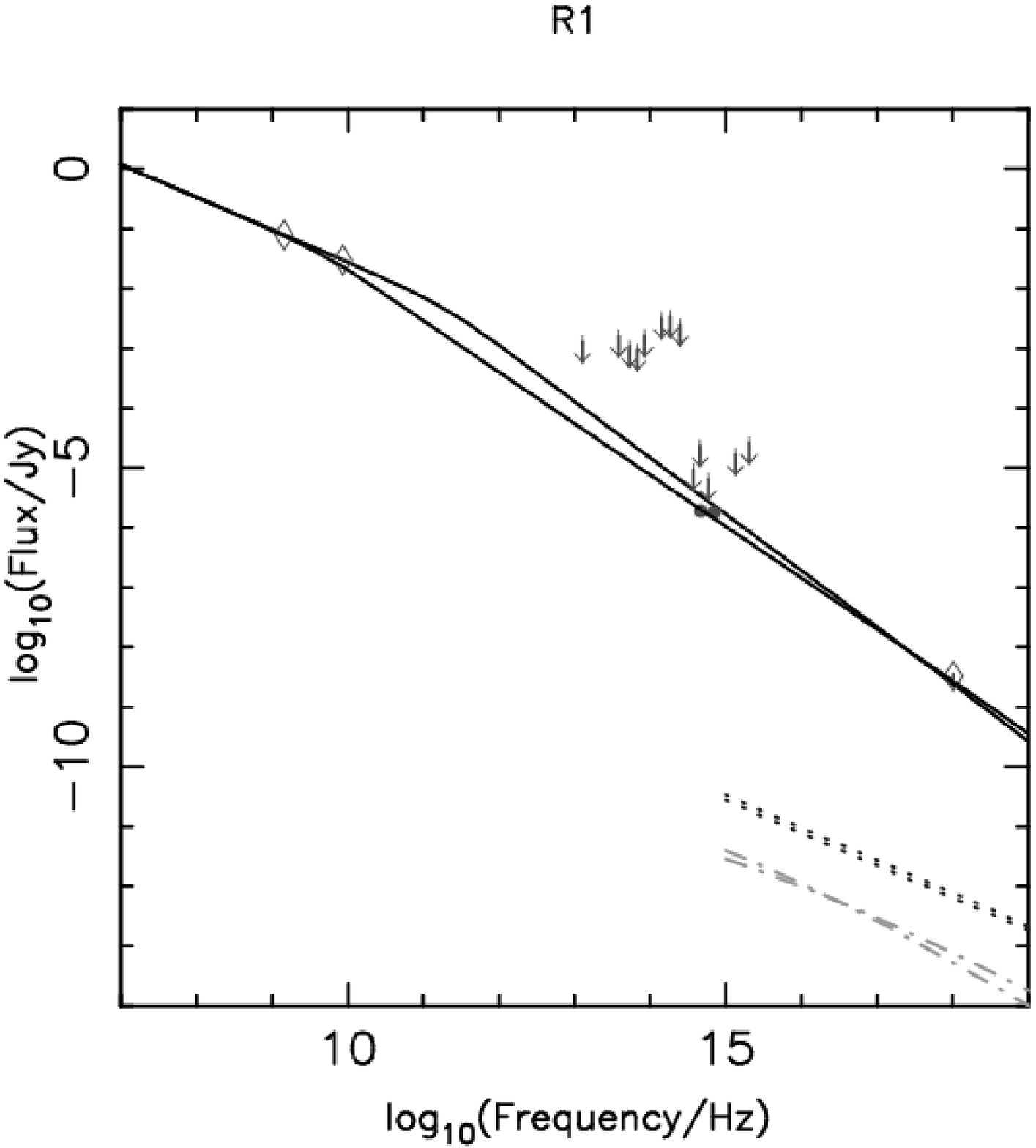}}
\caption{Spectrum of the jet in region R1.
The diamonds denote the flux densities of the detections, and the arrows
the 3$\sigma$ upper limits for the non-detections.  The circles are estimates
based on the measurements of \citet{cro03}. Two representative
models of the synchrotron output from broken power-law electron energy distributions
are plotted; one in which the break frequency
lies just above the 8.4 GHz radio point, and a second in which the break frequency
has the largest allowable value given the HST constraints.
The predicted IC/CMB and SSC X-ray flux densities assuming equipartition
are shown as the dotted and dash-dotted lines, respectively.
\label{sedr1}}
\end{figure}

\begin{figure}
\centerline{\includegraphics[width=0.9\linewidth]{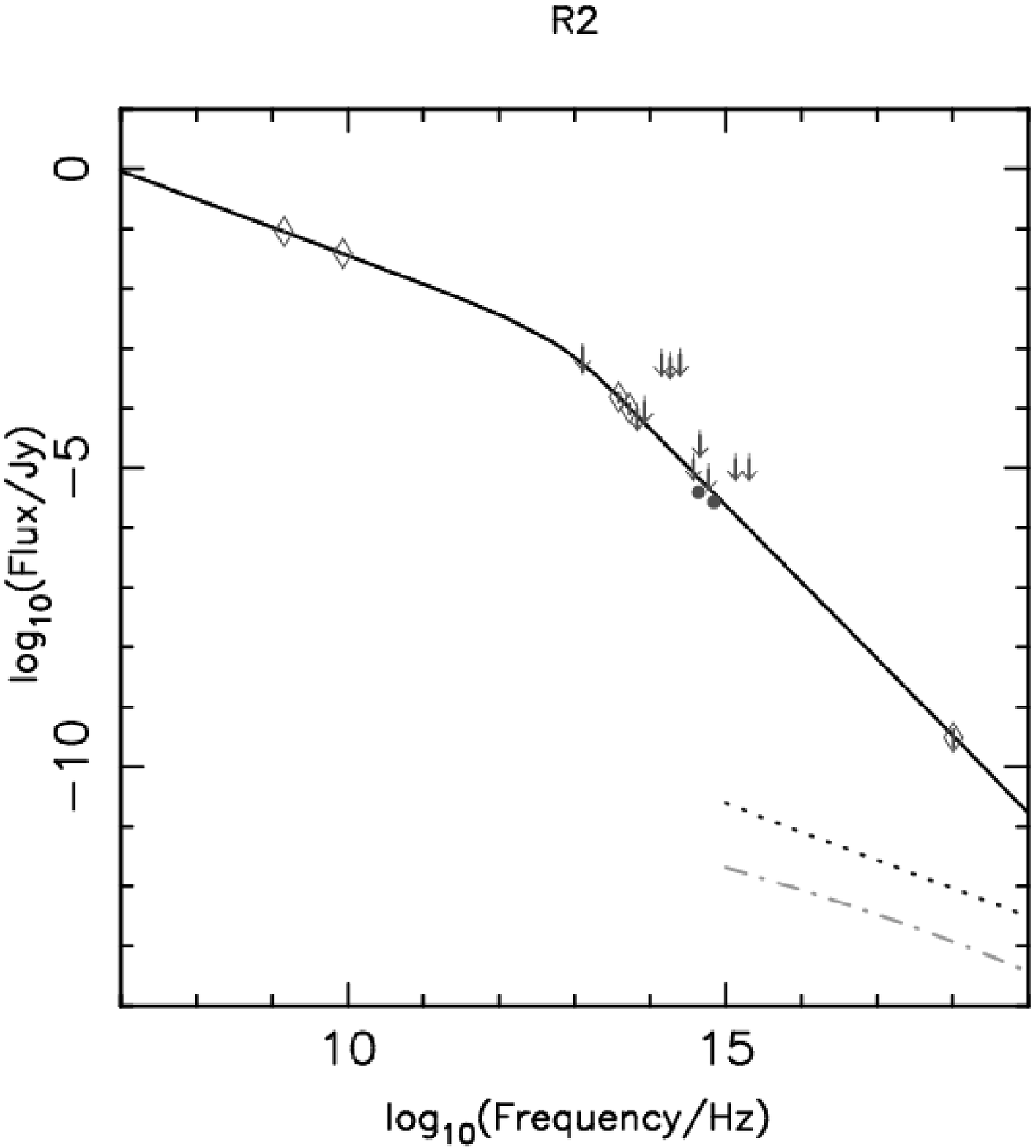}}
\caption{Spectrum of the jet in region R2.
The diamonds denote the flux densities of the detections, and the arrows
the 3$\sigma$ upper limits for the non-detections. The circles are estimates
based on the measurements of \citet{cro03}. The best fit
synchrotron model from a broken power-law electron energy distribution is also plotted. 
The predicted IC/CMB and SSC X-ray flux densities assuming equipartition
are shown as the dotted and dash-dotted lines, respectively.
\label{sedr2}}
\end{figure}

\begin{figure}
\centerline{\includegraphics[width=0.9\linewidth]{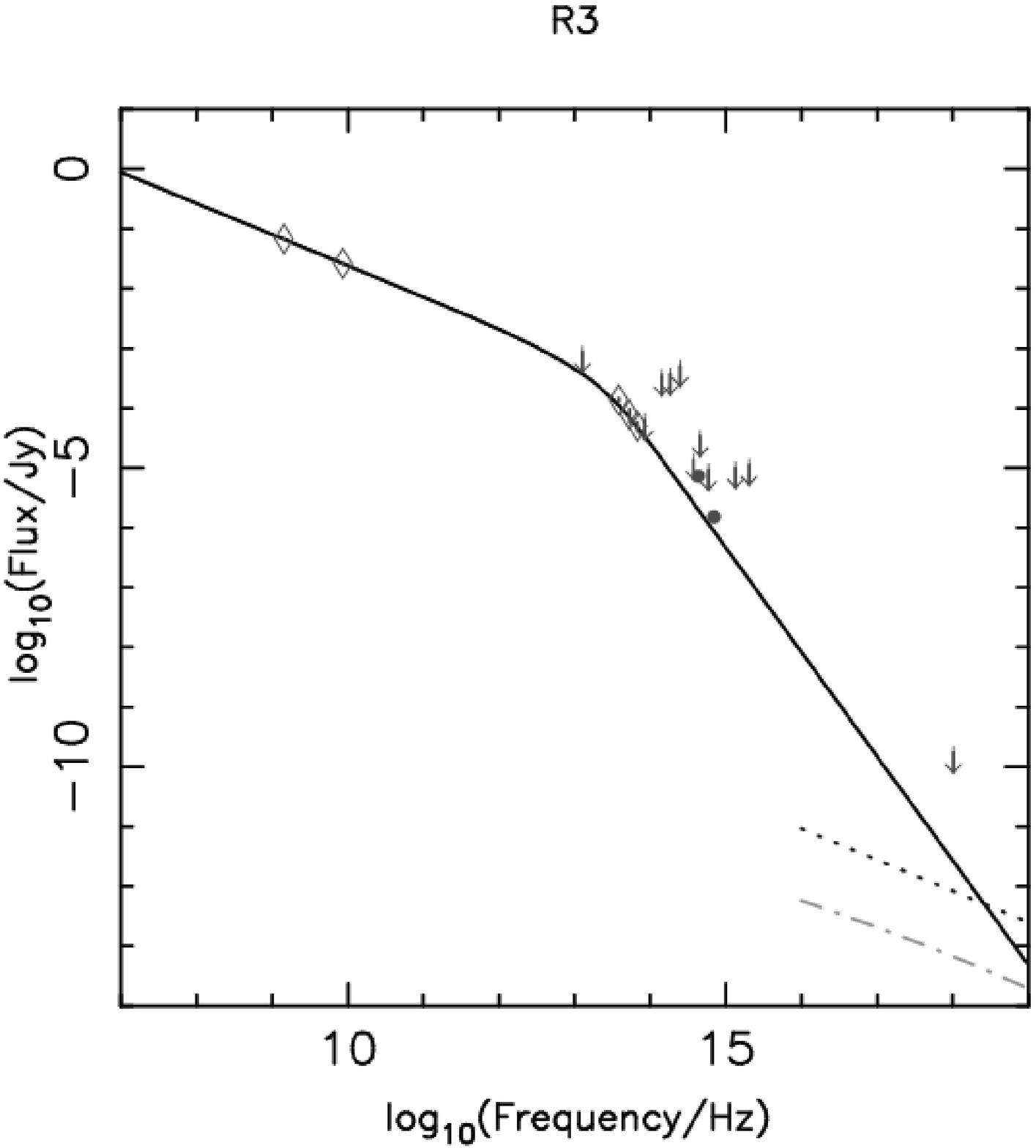}}
\caption{Spectrum of the jet in region R3.
The diamonds denote the flux densities of the detections, and the arrows
the 3$\sigma$ upper limits for the non-detections. The circles are estimates
based on the measurements of \citet{cro03}. The best fit
synchrotron model from a broken power-law electron energy distribution is also plotted. 
The predicted IC/CMB and SSC X-ray flux densities assuming equipartition
are shown as the dotted and dash-dotted lines, respectively.
\label{sedr3}}
\end{figure}

We measured the radio, infrared, and X-ray flux densities from
the jet in three different regions (labeled R1, R2, and R3; see Figure \ref{multiaps}) along the jet axis and 
set upper limits on the UV and optical fluxes in these regions.  Our choice of regions 
was guided by the morphology of the jet in the IR and by its relationship to the
X-ray emission seen with Chandra \citep{har02}.  
In the dynamical study of the 3C 31
jet by L02, the jet was divided into three regions based on its radio morphology and
modeled jet dynamics.  These three regions were termed the inner, the flaring, and the outer regions, and 
differ somewhat from our regions.
Region R1 contains the peak X-ray emission and extends from $2\farcs5$ 
(1.1~kpc) to $8\farcs3$ (3.6~kpc) from the nucleus with a 
width of 4$\arcsec$ (1.7~kpc).   This roughly corresponds to the flaring region
in the analysis of L02.  We divided the outer region 
as defined by L02 into two distinct regions, which we call R2 and R3. 
The R2 region extends from $8\farcs3$ (3.6~kpc) 
to $18\farcs3$ (7.9~kpc) from the nucleus, and the R3 region extends 
from $18\farcs3$ (7.9~kpc) to $28\farcs3$ (12.2~kpc). 
Both regions are $5\arcsec$ (2.2~kpc) in width. The inner region of L02, 
where the jet has roughly constant opening angle, is within
$2\farcs5$ of the nucleus and is not resolved from the host galaxy 
in the Spitzer data, so we did not use it in our analysis. 

In addition to the radio, IR, and X-ray detections of the jet,
UV, optical, and additional IR data points are taken from archival
GALEX, HST, and 2MASS and MIPS observations, respectively.
The fluxes with uncertainties and upper limits at all frequencies 
for the three regions are given in Table \ref{tflux}. We measured 
consistent fluxes in the IRAC bands from the nonstellar images 
resulting from the subtraction of the IRAF ELLIPSE 
galaxy model and of the 3.6\um~stellar emission model.

The broadband spectra from R1, R2, and R3 are shown in 
Figures~\ref{sedr1} through~\ref{sedr3}.
To better constrain the electron energy distributions 
(EEDs) in the three regions and quantify differences between
them, we fit the radio to X-ray spectra with power-law models.  
In all the model fits, we used only the actual detections with the constraint that the
fit model could not exceed any of the upper limits.
The limited number of points in the spectra and the relative simplicity of the models
does not warrant a more sophisticated statistical treatment of the upper limits.
Only some of the upper limits put strong constraints on the model fits.
We first fit single power-law models to the three spectra and found
that this simple model could be rejected at $>$99.9\% significance in all cases.

We therefore fit the spectra with models of synchrotron emission from broken
power-law electron energy distributions which contain four free parameters:
two spectral indices, a break frequency, and a normalization.
An acceptable fit is found for the spectra from all three regions with this model.
The model is overdetermined in the R1 region since we have only three data points
and a model with four free parameters.  In this region,
we consider two models with the break frequency frozen.
The optical upper limit from HST forces the break frequency to lie in the optical
or at lower frequencies, so we consider two models that bracket the extreme limits - 
one model (F1) with a break frequency just above the 8.4 GHz 
radio data point and a second model (F2) with a break frequency in the mid-IR, the
highest frequency that fits through the radio and X-ray detections and lies below
the optical upper limit.  
The best fit parameters for each of our three regions are summarized in Table~\ref{tfit}, and the best fit
emission models have been overplotted on Figures~\ref{sedr1} to~\ref{sedr3}. 

There are a few interesting points to note about the morphology of the jet in the radio, IR, and
X-ray bands.  X-ray emission is detected only in the R1 and R2 regions, the two
regions closest to the nucleus.  The presence of the X-ray emission suggests that there is on-going
particle acceleration in these regions, but that acceleration to the highest energies appears
to be less efficient in the outer parts of the jet.  Guided by our experience with 
Cen A \citep{kraft02,har03}, there is almost
certainly unresolved substructure in these two regions.  
This clearly demonstrates that the electron energy distributions, and probably the nature of the particle
acceleration in general, are distinctly different in each of these regions.
There are significant differences in the three spectra that clearly define three distinct regions of
particle acceleration.  The low-frequency (i.e. radio) spectral indices for the three regions are between 0.47 and 0.54,
typical values for radio plasma.  The break frequency in region R1 is poorly defined, but must lie somewhere
between 10-800 GHz.  The spectral index after the break is relatively flat, $\sim$0.86-0.97. 
Regions R2 and R3, in contrast, have fairly well defined break frequencies around 10$^{13}$ Hz, in the far-IR 
spectral regime, with a significant spectral steepening beyond the break.  In particular, the spectral index 
after the break is $\sim$1.28 and $\sim$1.72 in regions R2 and R3, respectively.

\begin{deluxetable*}{lcccr}
\tabletypesize{\scriptsize}
\tablecaption{Broken Power Law Fit Parameters\label{tfit}}
\tablewidth{0pt}
\tablehead{
\colhead{Region} & \colhead{Normalization} & \colhead{$\alpha$}
& \colhead{$\alpha$} &\colhead{Break Frequency}  \\
\colhead{} & \colhead{(Jy)} & \colhead{Low Frequency}
& \colhead{High Frequency} &\colhead{(GHz)}
}
\startdata
R1-F1\tablenotemark{1} & $6300 $ & $  0.54  $ &      $  0.86 $    &      $  8.4$\tablenotemark{2} \\
R1-F2\tablenotemark{1}& $6300 $  & $  0.54  $ &      $  0.97 $    &      $  780$\tablenotemark{2} \\
R2      & $1900$\tablenotemark{1}   & 0.47\tablenotemark{1} &      $  1.28\pm0.05 $    & $  5250\pm1330$ \\
R3      & $4200$\tablenotemark{1}   & 0.52\tablenotemark{1} &      $  1.72\pm0.12 $    & $ 18600\pm1500$ \\
\enddata
\tablecomments{Fit parameters obtained by $\chi^{2}$ minimization
using detected fluxes (see Table \ref{tflux} and requiring the upper limits to be respected.
Since the R1 model is overdetermined, we fix the break frequency at the two extremes
permitted by the upper limits and determine the remaining parameters. The uncertainties quoted
are the 90\% uncertainties.}
\tablenotetext{1}{The formal statistical uncertainties of these fits or parameters are negligible 
compared to other sources of uncertainty.}
\tablenotetext{2}{Extremal break frequencies set by the existing radio data and the HST upper limits.}
\end{deluxetable*}

Using the best-fit broken power-law models for the spectra, we
determined the equipartition magnetic fields and the break energies of the
electron distributions. 
Predictions for X-ray IC/CMB and
SSC emission have been computed using the {\it synch} program \citep{mjh98, mjh04}.
We assume that $\gamma_{min}$=10 and $\gamma_{max}$=10$^5$, $\kappa$=0 (the ratio of
non-radiating to radiating particles), that the radio plasma has unity filling factor, 
and that the emitting volumes can be modeled as cylinders in the
plane of the sky.  The corrections for bulk relativistic motion were small, even in R1 where
the jet velocity is greatest, due to the relatively large angle to the line of sight.
In the two regions where X-ray emission is detected, the observed X-ray flux density lies orders of magnitude above the
IC/CMB and SSC predictions, thus strengthening the conclusion of \citet{har02} that this is
synchrotron emission from a population of ultra-relativistic electrons.
We find equipartition magnetic fields of 3.0, 2.3, and 2.1 nT in regions R1, R2, and R3, respectively.
The equivalent magnetic pressures lie somewhat below the pressure of the hot ISM external 
to the jet \citep{lai02a}, and implies that there is additional pressure support 
inside the jet, perhaps from lower-energy relativistic electrons or relativistic or thermal protons.
At these magnetic field strengths, we find that the synchrotron lifetime for infrared 
emitting particles is $\sim3\times10^4$~yr.
In order for the observed X-ray flux densities to be the result of IC/CMB, the 
magnetic field would need to be two orders of magnitude below its equipartition
value.

The observed IR emission from the 3C 31 jet cannot be plausibly attributed to any
type of inverse-Compton or electron scattering process for two reasons.  
First, if the IR emission were not synchrotron radiation, the electron energy distribution 
would have to turn down dramatically in an equipartition field of 3 nT to explain the radio emission
but fall below the infrared emission ($\gamma\sim10^{5-6}$).  The detected X-ray
emission in regions R1 and R2 could not then be explained by synchrotron emission
from a simple extension of the electron energy distribution to $\gamma\sim$10$^8$.  As described 
above, IC/CMB and SSC emission can be ruled out, unless the jet is far from equipartition, so the 
origin of the X-ray emission would be unknown.  Second, the pressure and total energy of thermal 
electrons from stars in the jet required to create the IR emission from electron scattering would 
be unrealistically large.  In particular, the energy density of stellar photons at the position of 
the jet would be $\sim$5$\times$10$^{-14}$ ergs cm$^{-3}$,
either resulting in an unstable overpressured jet or requiring an unrealistically large energy flux
due to the required number of scatterers.
Alternatively, as an FR I radio galaxy, 3C 31 is a mis-directed BL Lac in the Unified Scheme of AGN
\citep{urr95}.
There could be a significant component of IR emission from the pc-scale jet that is not beamed in our
direction. In this scenario however, a typical BL Lac would require an electron density of $\sim$1 cm$^{-3}$
in its jet in order to account for the observed infrared flux. If 3C 31 were a very bright 
($\sim$2-3 orders of magnitude brighter) mis-directed BL Lac requiring a lower density in its 
jet to create the observed infrared flux, we would expect other morphological features not
seen in this system (e.g. the IR/optical beamed flux would be visible at large radii due to Thomson scattering
from the electrons in the hot ISM).

\section{Discussion}

We detected IR emission from two regions of the northern jet in 3C 31 within $\sim$ 15 kpc of the nucleus.
The third region R1 likely also contains IR emission, however, we cannot resolve it from the nucleus
and dust disk also present within this region. From the fitted spectra in R1, we estimate that IR 
emission is most likely approximately 10-30~$\mu$Jy.
The radio through IR/optical to X-ray spectra of these three regions of the 3C 31 jet
are very different, but each region mirrors in some manner features seen in other jets
(e.g. M87 \citep{shi07, for07}, Cen A \citep{har06}, 3C 66B \citep{tan00, har01}, and 3C 15 \citep{mar98}). 
The change in the jet from X-ray dominated to IR dominated further down the jet is also
seen in jets of greater power \citep[e.g. 3C 273; ][]{uch06}.
In the region closest to the nucleus, R1, the X-ray flux is the highest, the break frequency is
the lowest, and the spectrum is the
flattest after the break.  The relatively flat spectral index extending to the X-ray band
is similar to that seen in the knots of the inner kpc of the jet in Cen A \citep{har03}.
Further extending the analogy to the jet of Cen A, it is likely that this region of the 3C 31 jet contains significant
unresolved substructure in the X-ray and probably the optical/IR as well.

The IR emission in regions R2 and R3 is significantly enhanced relative to the X-ray in comparison
with region R1.  The break frequencies are also higher than in R1 (see Table~\ref{tfit}), as are the
post-break spectral indices.
The simplest explanation for the observed radio, IR, optical, and X-ray fluxes in the three regions
is that there is a single impulsive event in region R1, and the spectral steepening is
simply the result of progressive radiative losses.
The radiative lifetime of the relativistic particles emitting in the Spitzer band
is $\sim$3$\times$10$^4$ yrs, assuming they are radiating in the equipartition field.
If these particles are advected down the jet with velocity 0.5c, they would travel roughly
5 kpc.  The IR emission from the jet is detected to at least 10 kpc from the nucleus, so based simply
on this lifetime argument, it is plausible that the observed multi-frequency features of the jet
are the result of synchrotron aging.  However, the X-ray and IR morphology of the jet argues strongly
against this scenario.  In particular, the jet is clearly brightest in the IR near the boundary between
regions R2 and R3, and fainter closer to the nucleus within region R2.  This suggests that
(re-)acceleration is occurring near the boundary of R2/R3. The IR flux in region R1 is
not well constrained (we only infer it from the radio/optical/X-ray spectrum
due to contamination from the nucleus), but the progressive
brightening of the IR emission with increasing distance from the nucleus demonstrates that the jet
is being re-energized in regions R2/R3.
Additionally, there would be no X-ray synchrotron emission in region R2 in this model.
Given the radiative lifetimes, advection of ultra-relativisitic particles may play some role
in the appearance of the jet, but it clearly is not the whole story.

There must be a second site of particle acceleration
in regions R2/R3.  The dramatic change in spectral
shape between R1 and R2 suggests that the jet is much less efficient
at accelerating particles to $\gamma\sim$10$^{7-8}$ perhaps because it has significantly decelerated.  
Additionally, the morphological changes in the radio also suggest a significant change
in the flow dynamics from region R1 to R2/R3.
The calculations of L02 show a sharp drop in velocity between regions R1 and R2, and a 
more gradual deceleration of the jet between R2 and R3.  
The obvious question to ask is are the changes in particle acceleration efficiency
related to the morphological/dynamic changes in the jet and if so, how are they related.
More generally, the broad-band spectral features seen in 3C 31 are similar to that seen in a wide variety of radio galaxies.
The spectra of other extragalactic jets generally show a low frequency spectral index between 0.5 and 0.7 and 
are typically fit with a broken power law. 
The jets of 3C 66B, 3C 15, and the outer diffuse regions of the Cen A jet all have breaks
in their spectra around $10^{13}$~Hz and have high frequency spectral indices between 1.13 (Cen A, 3C 15)
and 1.35 (3C 66B), similar to region R2.
We speculate that the efficiency of ultra-relativistic
particle acceleration is a strong function of the jet velocity and/or the entrained mass, and
that this is a general feature of particle acceleration of all extragalactic jets on kpc scales.

Mass loading may have an important effect on the dynamics of particle acceleration.
In both Cen A and 3C 31, the jets are believed to be moderately relativistic over the region where X-ray and IR
synchrotron emission is detected.  In both cases, the mass swept up by the jet, either from embedded
AGB stars or from the ambient ICM, is relatively small. 
But in both cases, the IR synchrotron
emission extends considerably further from the nucleus than the X-ray emission.
There are many ways that a small amount of mass loading can fundamentally change the dynamics of the flow:  it
could change the internal sound speed, the equation of state, the Reynolds number, etc.
A full discussion of the effects of mass loading the jet on the particle acceleration is beyond the scope
of this paper, but given the similarities of 3C 31 and Cen A, the dynamics of these systems appear
to be qualitatively similar. More generally, deep radio observations of FR I jets typically
show knotted embedded structures \citep{wor07} similar to Cen A and 3C 31, suggesting that
dissipative processes are a common feature of FR I jets and play an important role in particle
acceleration.

The dramatic transition from the X-ray dominated to IR dominated jet
is most clear in Centaurus A where the X-ray jet narrows and disappears roughly 4 kpc from the nucleus,
while the IR emission brightens as the jet opens into the northeast radio lobe \citep{har06}.  
It is in this region where the jet dynamics appears to transition from momentum-dominated to pressure-dominated.
\citet{kraft08} showed that in the case of Cen A, the jet appears to cross a density/temperature discontinuity 
in the hot ISM.  
This discontinuity in density and temperature, but not pressure, is the result of
non-hydrostatic motions induced in the gas by a recent merger.  There are several galaxies
in the larger 3C 31 group, and one of these satellites may have disturbed
the gas core.  The transition from momentum to pressure domination in the jet may have an important
effect on the particle acceleration efficiency and hence the visible appearance of the jet at IR/optical/X-ray
wavelengths.  This suggests that subtle changes in the state of the external atmosphere
may have an important influence on the efficiency of electron acceleration internal to the jet.
The existing Chandra data are not sufficient to determine whether there is a surface brightness discontinuity
present in the gas at the interface between regions R1 and R2.
Alternatively, it has been argued that the X-ray synchrotron shocks in the Cen A jet are
the result of interactions between the jet and AGB stars in the host galaxy (Nulsen \etal, in preparation), and
that mass loading of the jet from material swept up via Kelvin-Helmholtz instabilities
plays little role in the deceleration of the jet.
There are some important differences between the Cen A jet \citep{kraft02} and that of 3C 31; 
the jet is somewhat smaller in Cen A and its synchrotron output is about an order of magnitude 
less powerful than in 3C 31, so one must be careful about pushing this analogy too far,
but there are clearly some qualitative similarities.

We obtained the external gas pressure measured as a function of distance from the nucleus
from \citet{lai02a}, and compared it to the equipartition pressure for each region.  In region R1,
the difference between the external pressure and the equipartition
pressure is nearly an order of magnitude, but in region R3, it is less than a factor of 2.
Note that the jet pressures cannot actually be this low, the jet would be crushed in an (ISM)
sound crossing time if this were true.  As argued by \citet{lai02a}, the trend that $P_{eq}$/$P_{gas}$ approaches unity
at larger distances from the nucleus suggests that mass entrainment and heating is not providing
the additional pressure support.  Relativistic, radiating particles and magnetic fields
are likely the dominant source of pressure support along the length of the jet. 
One would naively expect that more mass has been entrained at larger
radii and that if such entrainment played an important role in the pressure of the jet,
it would become progressively more important at larger radii.
That is, mass loading cannot play a significant role in providing pressure support, although as we have
argued above it likely plays a critical role in particle acceleration.  

Finally we note that the clear differences in radio flux density between
the jet and counterjet can be accounted for by relativistic beaming 
\citep{lai02a}, and the differences in the IR can be attributed to the same process.
We calculated the expected flux ratio in the jet and counterjet due to beaming:
\begin{equation}
R_{jet}=\left( \frac{1-\beta~cos(\theta)}{1+\beta~cos(\theta)} \right) ^{-(\alpha + 2)}
\end{equation}
Assuming an angle to the line of sight of $\theta=52^{\circ}$ and velocities 
from the models of L02 and using the appropriate spectral indices, we find
the expected ratio in the region where we measured the counterjet
flux to be $\sim$3 in the radio bands and $\sim$6 in the infrared bands. Our
measured ratios are close to these in the radio bands and the marginal
detections/upper limits on the counterjet's infrared flux yield lower limits on the 
flux ratio that agrees with the expected value. Since
the X-ray emission is strongest in R1, we searched for X-ray counterjet emission in a corresponding
region on the opposite side of the nucleus. We do not currently detect the counterjet. Using the
same velocity profile ($\beta$ between 0.72-0.76) and angle to the line of sight, we expect a 
ratio between 50 and 65 estimated over R1. The expected (0.5-2.0~keV) flux of $\sim0.1$~nJy
is lower than the upper limit of $0.6$~nJy currently set by the background in the existing Chandra
exposure.

\section{Conclusions}

We have presented a clear detection of IR emission from the northern jet and a marginal detection of the
counterjet of 3C 31.  We show that there is significant variation in the radio through
X-ray spectra along the jet, and have outlined some implications for the jet dynamics and particle
acceleration.  In particular, the ratio of IR to X-ray flux increases dramatically 
as a function of distance from the nucleus.
In a future paper, we will present results from Spitzer observations
of a sample of nearby radio jets (Bliss \etal, in preparation).
We are slowly increasing our sample size of IR, optical, and X-ray jets,
and it is now clear that broadband synchrotron emission is a ubiquitous feature
of FR I jets.  The spectral properties of the 3C 31 jet appear to span those seen
in other extragalactic jets.  It is likely that a range of particle acceleration regimes
are present in jets generally, and at least two different regimes are present in 3C 31.

Additional future observations may shed further light on some of the issues addressed
here.  In particular, a deep Chandra observation would be able to resolve the detailed
structure of the gas external to the jet.  Such an observation may reveal contact discontinuities
between two relatively stationary fluids that may play an important role in the jet dynamics.
A deeper X-ray observation of the jet may also reveal regions of lower surface brightness further
from the nucleus than observed here, as well as more clearly resolve the structures of the inner jet.
Such an observation also would put strong constraints on the presence of a counter jet and
constrain the jet velocity.
Higher angular resolution IR observations, particularly at lower frequencies than presented
here, may be able to resolve knots and other substructures in the jet.

\acknowledgements 
This work was based on archival data obtained
from the Spitzer Science Archive, the Chandra Data Archive, and the Multimission
Archive at STScI. This work was supported in part by the Smithsonian Institution,
the Chandra X-ray Center, NASA contract NNX07AQ18G, and the Harvard College Observatory.

{\it Facilities:} \facility{Spitzer}, \facility{CXO},  \facility{VLA},
	\facility{HST:WFPC2}, \facility{2MASS}, \facility{GALEX}


\begin{thebibliography}{}

\bibitem[Ashby \etal~(2009)]{ash09} Ashby, M.~L.~N., \etal~2009,  \apj, 701, 428
\bibitem[B\^{i}rzan \etal~(2004)]{bir04} B\^{i}rzan, L., Rafferty, D. A., 
	McNamara, B. R., Wise, M. W., \& Nulsen, P. E. J. 2004, \apj, 607, 800
\bibitem[Brookes \etal~(2006)]{bro06} Brookes, M.~H., Lawrence, C.~R., Keene, J., Stern, 
	D., Gorijan, V., Werner, M., \& Charmandaris, V. 2006, \apj, 646, L41
\bibitem[Burch(1977)]{bur77} Burch, S.~F. 1977, \mnras, 181, 599
\bibitem[Butcher \etal~(1980)]{but80} Butcher, H.~R., van Bruegel, W., 
	Miley, G.~K., 1980, \apj, 235, 749
\bibitem[Croston \etal~(2003)]{cro03} Croston, J.~H., Birkinshaw, M., Conway, E., \&
	Davies, R.~L. 2003, \mnras, 339, 82
\bibitem[de Vaucouleurs \etal~(1991)]{dev91} de Vaucouleurs, G., de
	Vaucouleurs, A., Corwin, J.~R., Buta, R.~J., Paturel, G., \&
	Fouque, P., 1991, Third Reference Catalogue of Bright Galaxies,
	(New York: Springer-Verlag)
\bibitem[Fanaroff \& Riley(1974)]{fan74} Fanaroff, B.~L. \& Riley, J.~M. 1974,
	\mnras, 167, 31P
\bibitem[Fazio \etal~(2004)]{faz04} Fazio, G.~G. \etal~2004, 
    \apjs, 154, 10
\bibitem[Fomalont \etal~(1980)]{fom80} Fomalont, E.~B., Bridle, A.~H., 
	Willis, A.~G., \& Perley, R.~A. 1980, \apj, 237, 418
\bibitem[Forman \etal~(2007)]{for07} Forman, W. \etal~2007, 
    \apj, 665, 1057
\bibitem[Hardcastle \etal~(1998)]{mjh98} Hardcastle, M. J., Birkinshaw, M., \& Worrall, D. M. 1998, \mnras, {\bf 294}, 615
\bibitem[Hardcastle \etal~(2001)]{har01} Hardcastle, M.~J., Birkinshaw, M., \& Worrall, D.~M., 2001, \mnras, {\bf 326}, 1499
\bibitem[Hardcastle \etal~(2002)]{har02} Hardcastle, M.~J., Worrall, D.~M., 
	Birkinshaw, M., Laing, R.~A., \& Bridle, A.~H. 2002, \mnras, 334, 182
\bibitem[Hardcastle \etal~(2003)]{har03} Hardcastle, M. J., Worrall, D. M., Kraft, R. P., Forman, W. R., Jones, C., \& Murray, S. S. 2003, \apj, {\bf 593}, 169
\bibitem[Hardcastle \etal~(2004)]{mjh04} Hardcastle, M. J., Harris, D. E., Worrall, D. M., \& Birkinshaw, M. 2004, \apj, {\bf 612}, 729
\bibitem[Hardcastle \etal~(2006)]{har06} Hardcastle, M.~J., Kraft, R.~P., \& 
	Worrall, D.~M. 2006, \mnras, 368, L15
\bibitem[Holtzman \etal~(1995)]{hol95} Holtzman, J.~A., Burrows, C. J., Casertano, S.,
	Hester, J.~J., Trauger, J.~T., Watzon, A.~M., \& Worthey, G. 1995, \pasp,
	107, 1065
\bibitem[Kraft \etal~(2002)]{kraft02} Kraft, R. P. \etal~2002, \apj, {\bf 569}, 54
\bibitem[Kraft \etal~(2008)]{kraft08} Kraft, R. P. \etal~2008, \apj, {\bf 677}, L97
\bibitem[Laing \& Bridle(2004)]{lai04} Laing, R.~A. \& Bridle, A.~H. 2004, \mnras, 348, 1459
\bibitem[Laing \& Bridle(2002a)]{lai02a} Laing, R.~A. \& Bridle, A.~H. 2002a, \mnras, 336, 1161
\bibitem[Laing \& Bridle(2002b)]{lai02b} Laing, R.~A. \& Bridle, A.~H. 2002b, \mnras, 336, 328 (L02)
\bibitem[Laing \etal~(2008)]{lai08} Laing, R.~A., Bridle, A.~H., Parma, P., Feretti, L.,
	Giovannini, G., Murgia, M., \& Perley, R.~A., 2008, \mnras, 386, 657
\bibitem[Lara \etal~(1997)]{lara97} Lara, L., Cotton, W. D., Giovannini, G., Venturi, T., \&
	Marcaide, J. M. 1997, \apj, 474, 179
\bibitem[Loken \etal~(1995)]{loken95} Loken, C., Roettiger, K., Burns, J. O., \& Norman M.
	1995, \apj, 445, 80
\bibitem[Martel \etal~(1998)]{mar98} Martel, A.~R. \etal~1998, \apj, 496, 203
\bibitem[Martel \etal~(2000)]{mar00} Martel, A.~R., Turner, N.~J., Sparks, W.~B., \&
	Baum, S.~A. 2000, \apjs, 130, 267
\bibitem[Martin \etal~(2005)]{mar05} Martin, D.~C. \etal~2005, \apj, 619, L1
\bibitem[Pahre \etal~(2004)]{pah04} Pahre, M.~A., Ashby, M.~L.~N., Fazio, G.~G.,
	\& Willner, S.~P. 2004, \apj, 154, 229
\bibitem[Peng \etal~(2002)]{pen02} Peng, C.~Y., Ho, L.~C., 
    Impey, C.~D., \& Rix, H. 2002, \aj, 124, 266
\bibitem[Rieke \etal~(2004)]{rie04} Rieke, G.~H. \etal~2004, \apjs, 154, 25
\bibitem[Schuster \etal~(2006)]{sch06} Schuster, M.~T., 
    Marengo, M. \& Patten, B.~M. 2006, SPIE, 6270, 65
\bibitem[Shi \etal~(2007)]{shi07} Shi, Y., Rieke, G.~H., Hines, D.~C., 
	Gordon, K.D., \& Egami, E. 2007, \apj, 655, 781
\bibitem[Tang \etal~(2009)]{tan09} Tang, Y., Gu, Q.-S., Huang, J.-S.,
	\& Wang, Y.-P. 2009, \mnras, 397, 1966
\bibitem[Tansley(1997)]{tan97} Tansley, D. 1997, M. S. Thesis, University of
	Manchester
\bibitem[Tansley \etal~(2000)]{tan00} Tansley, D., Birkinshaw, M., Hardcastle, M.~J., 
	\& Worrall, D.~M., 2001, \mnras, {\bf 317}, 623
\bibitem[Uchiyama \etal~(2006)]{uch06} Uchiyama, Y. et al. 2006, \apj, 648, 910
\bibitem[Urry \& Padovani(1995)]{urr95} Urry, M. \& Padovani, P. 1995, \pasp, 107, 803
\bibitem[Weisskopf \etal~(2000)]{wei00} Weisskopf, M.~C., Tananbaum, H.~D., Van Speybroeck, L.~P.,
	\& O'Dell, S.~L. 2000, SPIE, 4012, 2
\bibitem[Worrall(2009)]{wor09} Worrall, D.~M. 2009, A\&ARv 17, 1 
\bibitem[Worrall et al. (2007)]{wor07} Worrall, D.~M., Birkinshaw, M., Laing, R. A., Cotton, W.D.,
	\& Bridle, A.H. 2007, \mnras, 380, 2 
\end{thebibliography}
\end{document}